\documentclass[english,prl, twocolumn, footinbib, showpacs, showkeys]{revtex4}
\usepackage[T1]{fontenc}
\usepackage{color}
\usepackage{babel}
\usepackage{amsmath}
\usepackage{amssymb}
\usepackage{graphicx}
\usepackage[unicode=true,
 bookmarks=true,bookmarksnumbered=false,bookmarksopen=false,
 breaklinks=true,pdfborder={0 0 1},backref=false,colorlinks=true]
 {hyperref}
\hypersetup{pdftitle={Magneto-electric coupling in a two-dimensional ballistic Josephson junction with in-plane magnetic texture},
 pdfauthor={François Konschelle},
 pdfsubject={mesoscopic physics ; condensed-matter ; Josephson effect ; },
 pdfkeywords={Keywords: magneto-electric coupling ; \varphi -Josephson-junction ; anomalous current-phase relation ; spin-orbit ; magnetic texture ; ballistic transport equation ; chiral propagation ; two-dimensional gas ;},
 citecolor=blue, linkcolor=magenta,urlcolor=blue}
\usepackage{breakurl}

\DeclareGraphicsExtensions{.pdf}

\makeatletter
\@ifundefined{textcolor}{}
{%
 \definecolor{BLACK}{gray}{0}
 \definecolor{WHITE}{gray}{1}
 \definecolor{RED}{rgb}{1,0,0}
 \definecolor{GREEN}{rgb}{0,1,0}
 \definecolor{BLUE}{rgb}{0,0,1}
 \definecolor{CYAN}{cmyk}{1,0,0,0}
 \definecolor{MAGENTA}{cmyk}{0,1,0,0}
 \definecolor{YELLOW}{cmyk}{0,0,1,0}
}

\DeclareMathOperator{\Tr}{Tr}

\makeatother

\begin{document}

\title{Magneto-electric coupling in a two-dimensional ballistic Josephson
junction with in-plane magnetic texture}

\author{Fran\c{c}ois Konschelle}

\affiliation{JARA-Institute for Quantum Information, RWTH Aachen University, D-52074
Aachen, Germany}

\email{konschelle@physik.rwth-aachen.de}

\begin{abstract}
We study a Josephson junction made with a spin-textured bridge, when
both Rashba and Zeeman interactions combine to generate a magneto-electric
coupling between the superconducting current and the in-plane magnetic
texture in the normal region. In particular, we unambiguously obtain
the so-called anomalous current-phase relation $j=j_{c}\sin\varphi+j_{a}\cos\varphi$
in a two-dimensional ballistic Josephson junction close to the critical
temperature of the heterostructure, when an anomalous current $j_{a}\neq0$
subsists even at zero phase-difference between the superconductors,
and is responsible for the coupling between the magnetic and electric
degrees of freedom of the junction. The anomalous magneto-electric
current is due to the combination of the chirality of the propagating
modes and the anisotropy of the in-plane magnetic texture.
\end{abstract}



\date{\today}

\maketitle
Thanks to the recent progresses in heterostructures fabrication, it
becomes possible to study the Josephson effect in the presence of
strong spin-orbit effects \cite{Doh2005,Xiang2006,Nilsson2012}. Then
we can envision possible developments of spin-textured (ST) Josephson
systems, when the spin-orbit effect combines with the well-established
superconducting-ferromagnet (S/F) proximity effect \cite{buzdin.2005_RMP,bergeret_volkov_efetov_R.2005}.
This will certainly open the way to interesting discoveries, ranging
from the fundamental question of the macroscopic quantum effects under
exotic conditions \cite{casalbuoni.nardulli.2004,Alford2008}, to
topological order useful for quantum information \cite{Nayak2008},
passing through the coherent manipulation of mesoscopic circuits \cite{Georgescu2014}.

As an example of actual interest, the anomalous current-phase relation
$j=j_{c}\sin\varphi+j_{a}\cos\varphi$ (\textit{i.e.} $j=j_{0}\sin\left(\varphi+\varphi_{0}\right)$)
attracted some attentions over the past years \cite{Mints1998,buzdin.2005,Sickinger2012,krive_kadigrobov_shekhter_jonson.2005,buzdin:107005.2008,reynoso_etal:107001.2008,Zazunov2009,Liu2010,Reynoso2012,Yokoyama2012,Yokoyama2014}.
Such $\varphi_{0}$-junctions exhibit a non vanishing super-current
even at zero phase-difference between the two superconductors $j\left(\varphi=0\right)\neq0$,
called an anomalous current $j_{a}$. Their possible applications
are numerous: they produce a self-sustained flux when embedded in
a SQUID geometry \cite{Mints1998}, they can act as some phase batteries
in coherent circuits \cite{buzdin.2005}, they present a current asymmetry
and act as a supercurrent rectifier \cite{reynoso_etal:107001.2008},
... 

A Josephson junction (JJ) with an anomalous current necessarily breaks
the time-reversal symmetry since then $j\left(\varphi\right)\neq-j\left(-\varphi\right)$
\cite{golubov_kupriyanov.2004}. Then it seems natural to look for
a $\varphi_{0}$-junction in S/F/S systems \cite{buzdin.2005_RMP,bergeret_volkov_efetov_R.2005}.
So far, single junctions produced only $\varphi_{0}=\left\{ 0,\pi\right\} $,
the so-called $\pi$-JJ. In contrary, a parallel ($0$-$\pi$)-JJ
demonstrated the $\varphi_{0}$-behavior, with an extrinsic anomalous
current induced by an external magnetic field \cite[and references therein]{Sickinger2012}.
In a try to obtain an intrinsic anomalous current the spin-orbit interaction
enters the stage: since it allows a manipulation of the critical current
in two-dimensional JJ \cite{Takayanagi1985,Kresin1986}, the hope
is to obtain a JJ with a $\varphi_{0}$-phase-shift adjustable by
an external gate voltage. Then some JJs with both Zeeman and Rashba
interactions have been intensively investigated over the past few
years \cite[and references therein]{krive_kadigrobov_shekhter_jonson.2005,buzdin:107005.2008,reynoso_etal:107001.2008,Zazunov2009,Liu2010,Reynoso2012,Yokoyama2012,Yokoyama2014}.
\begin{figure}[b]
\includegraphics[width=0.95\columnwidth]{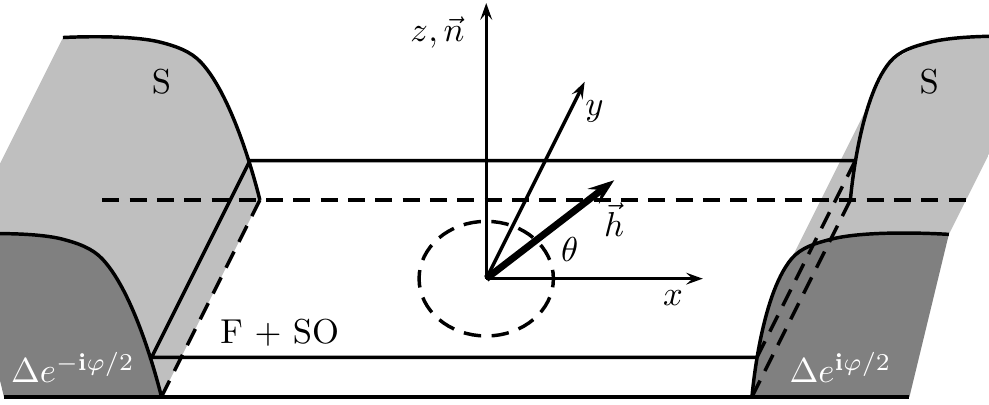}

\caption{Schematic of the junction studied in this paper. A two-dimensional
gas where the spin-orbit (along the $z$-axis, represented by $\boldsymbol{n}$)
and paramagnetic (in the $xy$-plane, represented by $\boldsymbol{h}$
and angle $\theta$) interactions compete (white region) is sandwiched
between two superconducting banks (gray regions). The phase difference
between the two superconducting electrodes is $\varphi$, and the
gap parameters $\Delta$ are constant and equals in the superconductors.
Note that the exchange interaction could be either due to a magnetic
interaction inside the junction, or to an applied magnetic field in
the $xy$-plane.\label{fig:schema}}
\end{figure}

In particular, Buzdin proposed a simple model for a S/ST/S Josephson
junction and found a remarkable expression for the phase-shift $\varphi_{0}\propto\boldsymbol{\left(h\times n\right)\cdot\nabla\varphi}$,
allowing a direct coupling between magnetization and super-current
in the junction \cite{buzdin:107005.2008}. Such a coupling was found
using symmetry arguments (see \textit{e.g.} \cite{Edelstein1996}
for a detailed derivation of the associated Ginzburg-Landau functional)
and can intrinsically be tuned by adjusting the paramagnetic interaction
$\boldsymbol{h}$ and/or the gate voltage, modifying the spin-orbit
vector $\boldsymbol{n}$. Then the super-current allows the generation
of the magnetization dynamics through the gradient of the superconducting
phase $\boldsymbol{\nabla\varphi}$ (the direction of the junction,
say) and back-action effect as well \cite{Konschelle2009}.

Despite its interesting consequences, the magneto-electric origin
of the anomalous current in ST-JJ has never been unambiguously clarified
in numerical analysis \cite{reynoso_etal:107001.2008,Reynoso2012,Zazunov2009,Liu2010,Yokoyama2012,Yokoyama2014}.
In particular, most of the simple models built-on from the numerical
investigations necessitate a mixing of chiral mesoscopic channels
for the derivation of an anomalous current $j_{a}$ \cite{reynoso_etal:107001.2008,Zazunov2009,Yokoyama2012,Yokoyama2014},
hindering its magneto-electric origins.

Here, we discuss the presence of magneto-electric effect in Josephson
physics. We use our recently proposed gauge-covariant transport formalism
\cite{Bergeret2014,Konschelle2014} to exhibit the anomalous $j=j_{c}\sin\varphi+j_{a}\cos\varphi$
current-phase relation in a S/ST/S-JJ where the normal region combines
Zeeman and Rashba interactions, with $j_{a}$ in \eqref{eq:ja} and
\eqref{eq:ja-ST}. We establish some generic expressions for $j_{c}$
and $j_{a}$ in \eqref{eq:jc} and \eqref{eq:ja} for a ballistic
2D-JJ in the presence of a non-trivial spin-field. This gauge-field
accounts for the chirality induced by the spin-orbit interaction.
In addition, the anisotropy of the Fermi surface, due to a paramagnetic
effect in the plane of the junction, leads to a geometric coupling
between the chiral propagation and the direction of the magnetic interaction.
We then show that these two minimal conditions (anisotropic chirality
and time-reversal-symmetry breaking) lead to a geometric coupling
$\varphi_{0}\propto\boldsymbol{\left(h\times n\right)\cdot\nabla\varphi}$
(see \eqref{eq:phi-0}), independent of the microscopic details:
those are not relevant in our quasi-classic formalism indeed.

We first discuss some of the difficulties to obtain simple solutions
of the complicated problem of a ST-JJ. For instance, if we suppose
a free-electron gas with spin-orbit, ferromagnetic and two-body BCS
interaction in a 1D configuration, there is no effect associated with
the spin-texture. To see this in more details, we write
\[
H=\int dx\left[\Psi^{\dagger}H_{\text{0}}\Psi+\dfrac{V\left(x\right)}{2}\left(\Psi^{t}\left(\mathbf{i}\sigma_{2}\right)\Psi\right)^{\dagger}\left(\Psi^{t}\left(\mathbf{i}\sigma_{2}\right)\Psi\right)\right]
\]
\begin{equation}
H_{0}=\dfrac{p_{x}^{2}}{2m}+\dfrac{p_{y}^{2}}{2m}-\mu+\boldsymbol{h\cdot\sigma}+v_{\text{so}}\boldsymbol{\left(n\times p\right)\cdot\sigma}\label{eq:H}
\end{equation}
for the Hamiltonian of the system, with $\Psi^{t}=\left(\begin{array}{cc}
\Psi_{\uparrow} & \Psi_{\downarrow}\end{array}\right)$ a spinor of fermionic annihilation operators, $\sigma_{i}$ the Pauli
matrices in the spin-space, $p_{x,y}$ the electron momentum in the
$x,y$-direction, $m$ the electron (effective) mass, $\mu$ the chemical
potential, $\boldsymbol{h}=h\left(\cos\theta,\sin\theta,0\right)$
the exchange field in the $xy$-plane, $v_{\text{so}}$ the Rashba
interaction strength of direction $\boldsymbol{n}$ (a unit vector
along the $z$-axis), and $V\left(x\right)$ the strength of the two-body
interaction, giving rise to superconductivity in some regions of space.
Note in passing that the Hamiltonian \eqref{eq:H} somehow describes
the dual setup than for Majorana modes (see \textit{e.g.} \cite[and references therein]{Cheng2012a,Rokhinson2012}
for some studies of the Josephson effect in the Majorana geometry),
since in the later case the Zeeman interaction is along the $z$-axis,
whereas we choose $\boldsymbol{h}$ to lie in the $xy$-plane instead.

Now we reduce the problem to 1D, say the $x$-axis of the junction,
and we choose $\theta=\pi/2$ in order to maximize $\varphi_{0}$.
Then, neglecting the $p_{y}$ component, one can show that the transformation
$\Psi\rightarrow R\Psi$ with $R=\exp\left[-\mathbf{i}\sigma_{2}\int mv_{\text{so}}dx/\hbar\right]$
removes the spin-orbit interaction without affecting the singlet-pairing
term in the BCS Hamiltonian. Thus under the above hypothesis, the
model never exhibits any anomalous current; of course the model can
still be used to study all the phenomenology associated with the S/F
proximity effect \cite{buzdin.2005_RMP,bergeret_volkov_efetov_R.2005}. 

So in order to establish a magneto-electric coupling, we have to discard
one of our hypothesis: either the junction should be explicitly two-dimensional,
or the band dispersion should not be quadratic. Buzdin already discussed
the second option \cite{buzdin:107005.2008}: when the dispersion
is not quadratic, the ST might not be gauge trivial, even in 1D problems.
Most of the numerical works in contrary discussed the first option
; then, in order for some anomalous current to exist when the band
dispersion is quadratic, a pinch-off of the structure seems to be
necessary \cite{reynoso_etal:107001.2008,Zazunov2009,Yokoyama2012,Yokoyama2014}.
Indeed, the anomalous current-phase relation is picturesquely supposed
to come from the mixing of different channels along the junction in
these studies. This additional difficulty renders difficult the establishment
of the geometric structure of the anomalous current $j_{a}$.

In the following, we study a ST-JJ, when the normal part consists
in a two-dimensional material lying in the $xy$-plane in the region
$-L/2\leq x\leq L/2$, with both Zeeman and Rashba interactions, and
sandwiched between two conventional ($s$-wave) superconductors in
the regions $x\leq-L/2$ and $x\geq L/2$, see Fig. \ref{fig:schema}.
Following the approach in \cite{Bergeret2014,Konschelle2014}, the
gauge-covariant transport equation associated with the Hamiltonian
\eqref{eq:H} reads, in the clean limit
\begin{equation}
-\mathbf{i}\hslash\left(\boldsymbol{v_{F}\cdot\nabla}\right)\mathbf{g}+\left[\mathbf{h}+\boldsymbol{\Delta},\mathbf{g}\right]-\mathbf{i}mv_{\text{so}}^{2}\partial_{\phi}\left\{ \sigma_{3},\mathbf{g}\right\} =0\label{eq:transport}
\end{equation}
where $\boldsymbol{v_{F}}=v_{F}\left(\cos\phi,\sin\phi\right)$ is
the projection of the Fermi velocity on the $\left(x,y\right)$-directions,
respectively. We use the parameterisation
\begin{equation}
\mathbf{g}=\left(\begin{array}{cc}
g & -f\\
f^{\dagger} & g^{\dagger}
\end{array}\right)\;;\; g=\left(\begin{array}{cc}
g_{\uparrow} & g_{+}\\
g_{-} & g_{\downarrow}
\end{array}\right)\label{eq:g-parameterisation}
\end{equation}
in the particle-hole and spin spaces, respectively, and the same parameterisation
is used for the $f$, $f^{\dagger}$ and $g^{\dagger}$ sectors. Also,
$\mathbf{h}=\mathbf{h}_{0}+\mathbf{h}_{Z}+\mathbf{h}_{A}$ where 
\begin{align}
\mathbf{h}_{0} & =\hbar\omega\tau_{3}\nonumber \\
\mathbf{h}_{Z} & =h\left(\cos\theta\sigma_{1}+\sin\theta\sigma_{2}\right)\tau_{3}=\left(\boldsymbol{h\cdot\sigma}\right)\tau_{3}\nonumber \\
\mathbf{h}_{A} & =p_{F}v_{\text{so}}\left(\sin\phi\sigma_{1}-\cos\phi\sigma_{2}\right)=h_{A}\hat{1}\label{eq:h}
\end{align}
represent the different interactions. The matrix $\boldsymbol{\Delta}=\Delta e^{\mathbf{i}\varphi}\tau_{+}-\Delta e^{-\mathbf{i}\varphi}\tau_{-}$
represents the gap parameter with $\varphi$ the superconducting phase-difference
across the junction, the $\tau_{i}$ are Pauli matrices describing
the particle-hole degree of freedom with $\tau_{\pm}=\left(\tau_{1}\pm\mathbf{i}\tau_{2}\right)/2$.
We further suppose that the gap parameters and the phases are constant
in the superconducting regions, and vanish in the normal regions.

\begin{figure}[b]
\begin{center}\includegraphics[width=0.75\columnwidth]{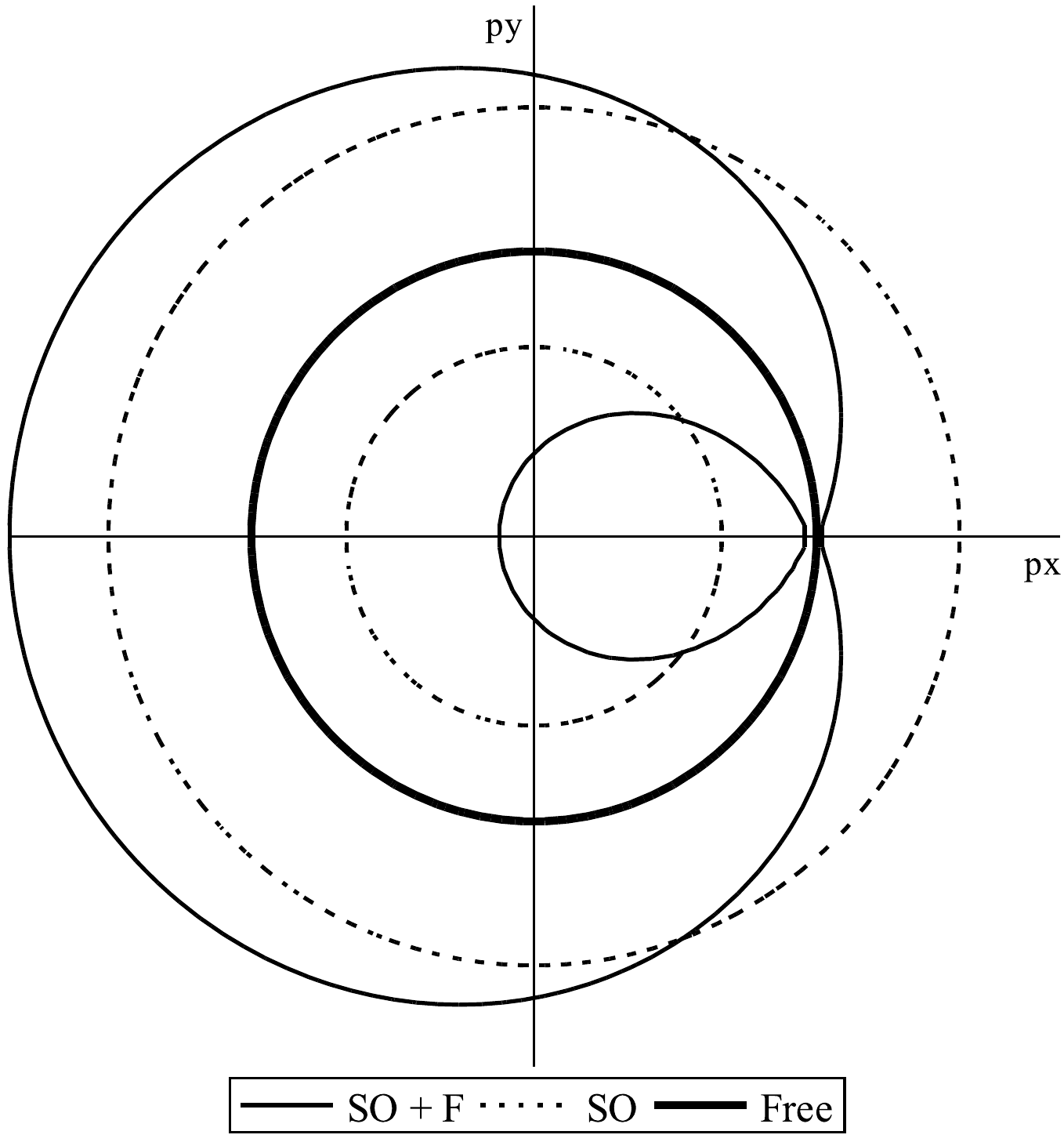}\end{center}

\caption{Fermi surfaces for non-interacting particles with and without magnetic
interaction, described by the Hamiltonian \eqref{eq:H}. A free
gas with a Rashba interaction leads to two spin-sub-bands (dashed
lines, labelled SO). The addition of the Zeeman interactions leads
to some anisotropy between these two surfaces (the two plain curves,
labelled SO + F). In contrary, the gauge-covariant quasi-classic method
simply requires a free-electron gas (\textit{i.e.} a quadratic dispersion
relation, the bold circle labelled Free) whereas the spin-orbit effect
is treated as a gauge-potential, leading to a non-trivial gauge-field
(see the text). Remarkably, the direction of maximum anisotropy is
given by $\boldsymbol{h\times n}$, see Eq.\eqref{eq:H}. In the plot,
the Zeeman interaction is supposed really strong and along the $y$-axis,
such that the anisotropy is mainly along the $x$-axis. The gauge-covariant
quasi-classic expansion in principle breaks long before reaching such
a large anisotropy and Fermi surface splitting, intended for illustration
purpose only.\label{fig:Fermi-surface}}
\end{figure}

We next recognize in \eqref{eq:h} the free-particle term $\mathbf{h}_{0}$,
the Zeeman $\mathbf{h}_{Z}$ and the spin-orbit $\mathbf{h}_{A}=\boldsymbol{v_{F}\cdot A}$
interactions, linearised in the proximity of the Fermi surface. More
importantly, the Fermi surface is supposed to be a single circle with
a constant radius $p_{F}$ in momentum space, parameterised by the
angle $\phi$ in \eqref{eq:transport}. Indeed, the main advantage
of the gauge-covariant formulation adapted to the transport formalism
is to consider a free-electron gas, and to discuss the spin-orbit
interaction as a non-Abelian gauge-potential $\boldsymbol{A}=mv_{\text{so}}\left(-\sigma_{y},\sigma_{x},0\right)/\hbar$
\cite{Bergeret2014,Konschelle2014}. It results a non-trivial gauge
field $F_{xy}=\partial_{x}A_{y}-\partial_{y}A_{x}+\mathbf{i}\left[A_{x},A_{y}\right]\propto mv_{\text{so}}^{2}\sigma_{3}$,
leading to the last term in \eqref{eq:transport}. This term takes
into account the difference in angular-momentum between the two spinful
sectors, see Fig. \ref{fig:Fermi-surface}, and thus it is responsible
for the chirality of the quasi-classical trajectories. An other important
ingredient of the model is the Zeeman interaction $\mathbf{h}_{Z}$
in \eqref{eq:h}. This interaction could be treated as a gauge-field
as well \cite{Konschelle2014}, but here we suppose instead that the
Fermi energy $E_{F}\gg h$, and we treat $\mathbf{h}_{Z}$ as a usual
potential in the quasi-classic approximation \cite{Bergeret2014}.
Due to the presence of the ST, the genuine Fermi surface given by
the spectrum of $H_{0}$ in \eqref{eq:H} is deformed, see Fig.
\ref{fig:Fermi-surface}. In our gauge-covariant transport formalism,
in contrary, the Fermi surface is always supposed circular, and the
anisotropy is taken into account as the combining effect of the chirality
in the presence of the magnetic interaction.

We want to establish the current density flowing through a two-dimensional
S/ST/S-JJ (see \textit{e.g.} \cite{Konschelle2014})
\begin{equation}
j=\mathbf{i}\pi eN_{0}v_{F}\int\dfrac{d\hbar\omega}{2\pi}\int_{0}^{2\pi}\dfrac{d\phi}{2\pi}\cos\phi\Tr\left\{ \tau_{3}\mathbf{g}\right\} \label{eq:current-clean-def}
\end{equation}
as a function of the exchange field, spin-orbit strength and phase
difference across the JJ. The current being conserved, we can evaluate
it at any position along the $x$-axis ; hereafter we choose the position
$x=0$ in the middle of the normal region.

One can show that the system of equations \eqref{eq:transport}
reduces to the form
\begin{equation}
\mathbf{i}\dfrac{d\mathbf{g}}{ds}=\dfrac{L}{\hbar v_{F}}\left[\mathbf{h}+\boldsymbol{\Delta},\mathbf{g}\left(s\right)\right]\label{eq:transport-Heisenberg}
\end{equation}
where  $\mathbf{g}\left(s\right)$ is a short-hand notation for $\mathbf{g}\left(z_{\zeta},w_{\zeta},\phi_{\zeta}\right)$
and with the characteristics
\begin{align}
\phi_{\zeta}\left(s\right) & =\phi+2\zeta s\nonumber \\
x_{\zeta}\left(s\right) & =-\dfrac{L}{2\zeta}\sin\phi+\dfrac{L}{2\zeta}\sin\left(2\zeta s+\phi\right)\nonumber \\
y_{\zeta}\left(s\right) & =\dfrac{L}{2\zeta}\cos\phi-\dfrac{L}{2\zeta}\cos\left(2\zeta s+\phi\right)\label{eq:parametric-characteristics}
\end{align}
with $\zeta$ a parameter due to the non-trivial gauge-field. This
important parameter exhibits the hallmark of the spin-orbit effect
in term of the quasi-classic trajectories. From \eqref{eq:transport}
we realize that the only components affected by the anti-commutator
are the spinful components $g_{\uparrow}$, $g_{\downarrow}$, $f_{\uparrow}$,
... The associated characteristics are circular trajectories characterized
by the parameter $\zeta=\pm\left(L/\xi_{\text{so}}\right)\left(v_{\text{so}}/v_{F}\right)$,
with $\xi_{\text{so}}=\hbar/mv_{\text{so}}$ a spin-orbit length.
So we see that, in addition to being bent, the trajectories of the
spinful components are also spin-polarised, reminiscent of the chiral
propagation of particles in systems with a strong spin-orbit interaction.
In contrary, the characteristic trajectories associated with the singlet
components $g_{+}$, $g_{-}$, $f_{+}$, ... are not affected by the
chiral symmetry, and thus propagate along straight lines $\zeta=0$
in \eqref{eq:parametric-characteristics}. Importantly, the limit
$\zeta\rightarrow0$ is not perturbative in \eqref{eq:transport-Heisenberg}:
a non-zero $\zeta$ modifies the topology of the trajectories, which
become circles for the spinful components. The strategy to obtain
the quasi-classic Green functions in \eqref{eq:transport-Heisenberg}
is thus clear: we first resolve the system \eqref{eq:transport-Heisenberg}
for a general $s$, and afterwards we substitute the characteristics
\eqref{eq:parametric-characteristics} according to
\begin{equation}
g=\left(\begin{array}{cc}
g_{\uparrow}\left(s_{\zeta=Lv_{\text{so}}/\xi_{\text{so}}v_{F}}\right) & g_{+}\left(s_{\zeta=0}\right)\\
g_{-}\left(s_{\zeta=0}\right) & g_{\downarrow}\left(s_{\zeta=-Lv_{\text{so}}/\xi_{\text{so}}v_{F}}\right)
\end{array}\right)\label{eq:g-s-zeta}
\end{equation}
and the same substitution for the other spin-matrices $f$, $f^{\dagger}$
and $g^{\dagger}$, where $s_{\zeta}$ is the inverse of the system
\eqref{eq:parametric-characteristics} depending parametrically
on $\zeta$ (see \eqref{eq:s-z} below).

Under the form \eqref{eq:transport-Heisenberg}, the transport
equation admits an Ansatz 
\begin{equation}
\mathbf{g}\left(s\right)=\mathbf{u}\left(s\right)\mathbf{g}_{0}\mathbf{u}^{-1}\left(s\right)+\mathbf{g}_{\infty}\label{eq:Ansatz}
\end{equation}
with $\mathbf{g}_{0}$ and $\mathbf{g}_{\infty}$ some constant matrices
\cite{Schopohl1995,Schopohl1998}. Such an Ansatz allows to resolve
Eq.\eqref{eq:transport-Heisenberg} in the form of the commutator
relation $\left[\mathbf{h}+\boldsymbol{\Delta},\mathbf{g}_{\infty}\right]=0$
in addition to the Schr\"{o}dinger-like equation
\begin{equation}
\mathbf{i}\dfrac{d\mathbf{u}}{ds}=\dfrac{L}{\hbar v_{F}}\left(\mathbf{h}+\boldsymbol{\Delta}\right)\mathbf{u}\left(s\right)\label{eq:transport-Schrodinger}
\end{equation}
solved directly in a matrix form for the propagation-like operator
$\mathbf{u}$ instead of resolving each components of the $\mathbf{g}$
matrix individually. Also, an expression like \eqref{eq:transport-Schrodinger}
can be perturbatively treated as usual with a Schr\"{o}dinger-type
equation \cite{Blanes2009}.

In the superconductors with a $s$-independent $\boldsymbol{\Delta}$,
and also $\mathbf{h}_{Z}=\mathbf{h}_{A}=0$, the gauge-field disappears,
so the trajectories are some straight lines given by $\zeta=0$ in
\eqref{eq:parametric-characteristics}. Injecting the Ansatz \eqref{eq:Ansatz}
into \eqref{eq:transport-Heisenberg}, one has $\left[\mathbf{h}_{0}+\boldsymbol{\Delta},\mathbf{g}_{\infty}\right]=0$
which has for solution $\mathbf{g}_{\infty}=N_{g}\left(\mathbf{h}_{0}+\boldsymbol{\Delta}\right)$
with the constant $N_{g}$ such that $\mathbf{g}_{\infty}^{2}=1$.
It is important to note that the normalization property is preserved
due to the absence of the gauge-field in the superconductor, see \textit{e.g.}
\cite{Konschelle2014}. Defining for commodity $\cos\eta=\hbar\omega/\Delta$
and $\xi=\hbar v_{F}/\Delta$ for the superconducting coherence length,
one has
\begin{align}
\mathbf{g}\left(x\leq-\dfrac{L}{2}\right) & =e^{-\mathbf{i}\tau_{3}\frac{\varphi}{4}}\left[\mathbf{S}_{+}g_{1}\tau_{+}\mathbf{S}_{+}^{-1}+\mathbf{g}_{\infty}^{0}\right]e^{\mathbf{i}\tau_{3}\frac{\varphi}{4}}\nonumber \\
\mathbf{g}\left(x\geq\dfrac{L}{2}\right) & =e^{\mathbf{i}\tau_{3}\frac{\varphi}{4}}\left[\mathbf{S}_{-}g_{2}\tau_{-}\mathbf{S}_{-}^{-1}+\mathbf{g}_{\infty}^{0}\right]e^{-\mathbf{i}\tau_{3}\frac{\varphi}{4}}\label{eq:sol-supra}
\end{align}
in the two superconductors, with $\mathbf{S}_{\pm}=\left(\tau_{\downarrow}-e^{\mathbf{i}\eta}\tau_{\uparrow}+\tau_{-}-e^{-\mathbf{i}\eta}\tau_{+}\right)e^{\tau_{3}\sin\eta\left(x\pm L/2\right)/\xi}$
and $\mathbf{g}_{\infty}^{0}=\mathbf{i}\left(\cos\eta\tau_{3}+\mathbf{i}\tau_{2}\right)/\sin\eta$,
with $\tau_{\uparrow,\downarrow}=\left(1\pm\tau_{3}\right)/2$. The
two matrices $g_{1,2}$ are now constant matrices in the spin-space
given by boundary conditions. In \eqref{eq:sol-supra} we see
why we needed the constant matrix $\mathbf{g}_{\infty}$: it accounts
for the bulk properties of superconductivity, on top of which some
evanescent waves are localized close to the interfaces. 

In the normal part, we suppose $\boldsymbol{\Delta}=0$, and thus
$\mathbf{g}_{\infty}=0$ in the Ansatz \eqref{eq:Ansatz} since
we expect to find propagating modes instead of evanescent ones. Still,
the term $\mathbf{h}_{A}$ is $s$-dependent, since it depends on
the angle $\phi\equiv\phi_{\zeta}\left(s\right)$. We have thus $\mathbf{g}\left(s\right)=\mathbf{u}\left(s\right)\mathbf{g}_{0}\mathbf{u}^{\dagger}\left(s\right)$
with the propagation operator
\begin{equation}
\mathbf{u}=e^{-\mathbf{i}\frac{\omega L}{v_{F}}s\tau_{3}}\left(\begin{array}{cc}
u_{+} & 0\\
0 & u_{-}
\end{array}\right)\label{eq:u-parameterisation}
\end{equation}
\begin{equation}
\mathbf{i}\dfrac{du_{\pm}}{ds}=\dfrac{L}{\hbar v_{F}}\left(\pm\left(\boldsymbol{h\cdot\sigma}\right)+h_{A}\right)u_{\pm}\left(s\right)\label{eq:u_pm}
\end{equation}
in the particle-hole space. We do not need an explicit and/or perturbative
form for $u_{\pm}\left(s\right)$ until Eq.\eqref{eq:u-perturb}
below.

Next step is to use rigid boundary conditions to obtain the generic
form of the particle-sector spin-matrix $g_{0}$ in the normal region.
Thanks to the decoupling between particles and holes in the normal
region (see \eqref{eq:u-parameterisation}), one can write the
formula
\begin{equation}
j=j_{c}\sin\varphi+j_{a}\cos\varphi\label{eq:CPR-clean}
\end{equation}
\begin{multline}
\dfrac{j_{c}}{j_{\Delta}}=-2\int_{-\pi/2}^{\pi/2}d\phi\cos\phi\times\\
\sum_{\Sigma}e^{\Sigma L\left(s_{L}^{\Sigma}-s_{R}^{\Sigma}\right)/2\xi_{T}}\Tr\left\{ U_{L}\cdot U_{R}^{\dagger}\right\} \label{eq:jc}
\end{multline}
\begin{multline}
\dfrac{j_{a}}{j_{\Delta}}=\dfrac{\mathbf{i}}{2}\int_{-\pi/2}^{\pi/2}d\phi\cos\phi\times\\
\sum_{\Sigma}e^{\Sigma L\left(s_{L}^{\Sigma}-s_{R}^{\Sigma}\right)/2\xi_{T}}\Tr\left[\left\{ U_{R},U_{L}^{\dagger}\right\} -\left\{ U_{R}^{\dagger},U_{L}\right\} \right]\label{eq:ja}
\end{multline}
close to the critical temperature $T_{c}$ of the junction, where
we defined $U=u_{+}^{\dagger}\left(s\right)u_{-}\left(s\right)$,
$U_{L,R}=U\left(s_{L,R}^{\Sigma}\right)$ and $j_{\Delta}=eN_{0}v_{F}\Delta^{2}/\pi^{2}k_{B}T_{c}$.
To find \eqref{eq:CPR-clean} we transformed the expressions for $\mathbf{g}$
both in the superconductors and in the normal region to the Matsubara
representation $\omega\rightarrow\mathbf{i}\omega_{n}$, $\hbar\omega_{n}\approx\pi k_{B}T_{c}\left(2n+1\right)$,
we expanded in the small $\Delta/k_{B}T_{c}$ parameter, and we supposed
$L/\xi_{T}\gg1$ with the thermal length $\xi_{T}=\hbar v_{F}/4\pi k_{B}T_{c}$.
We also defined $s_{L,R}^{\Sigma}=s\left(x=\mp L/2,\Sigma\right)$
as the expressions for the $s$-parameter at the boundaries between
the superconductors and the spin-textured regions and depending in
the direction of the propagation $\Sigma=\pm1$.

Eq.\eqref{eq:CPR-clean} is the main result of this letter. It gives
a generic expression for the first harmonics of the Josephson current
in a clean junction. Especially, it contains an explicit expression
for the anomalous current $j_{a}$. We see in the structure of the
$U_{L,R}$ matrix that it corresponds to the propagation of an electron
(hole) from $s=0$ (the middle of the junction in our parametrisation)
toward the point $s_{L,R}$ corresponding to the superconducting interface,
where it is reflected back as a hole (electron) to $s=0$. So the
product $U_{L}U_{R}^{\dagger}$ represents the entire exploration
of the particles inside the normal region, giving rise to the Andreev
modes in the junction which in turn transport the current in \eqref{eq:jc}.
In contrary, the anomalous current is due to some interferences between
the Andreev modes in \eqref{eq:ja}. Clearly, $j_{a}$ vanishes for
non magnetic bridges, since $u_{\pm}=1$ in that case: to break the
time-reversal symmetry is mandatory to obtain an anomalous current.

To go further one needs the expression of $s\left(x_{\zeta}\right)$,
which is given perturbatively as 
\begin{equation}
s\approx\Sigma\left(\dfrac{x/L}{\cos\phi}+\left(\dfrac{x}{L}\right)^{2}\dfrac{\sin\phi}{\cos^{3}\phi}\zeta+\cdots\right)\label{eq:s-z}
\end{equation}
from \eqref{eq:parametric-characteristics} in the limit $v_{\text{so}}\ll v_{F}$.
We see how the small chirality $\zeta=\left(0,\pm Lv_{\text{so}}/\xi_{\text{so}}v_{F}\right)$
affects the boundary conditions in the above expression. We then expand
the expressions \eqref{eq:jc} and \eqref{eq:ja} for a small spin-orbit
interaction. We obtain for instance ($j_{c}$ does not contain any
signature of the spin-orbit effect at first order)
\begin{multline}
\dfrac{j_{a}}{j_{\Delta}}=-\dfrac{L^{2}}{\xi_{\text{so}}\xi_{f}}\dfrac{v_{\text{so}}}{v_{F}}\int_{-\pi/2}^{\pi/2}d\phi\dfrac{\sin\phi}{\cos^{2}\phi}\times\\
e^{-L\tilde{s}/\xi_{T}}\Tr\left[\sigma_{3}\left\{ u_{+}^{\dagger}\left(\boldsymbol{\hat{h}\cdot\sigma}\right)u_{-},\bar{u}_{-}^{\dagger}\bar{u}_{+}\right\} \right.\\
\left.-\sigma_{3}\left\{ u_{+}^{\dagger}u_{-},\bar{u}_{-}^{\dagger}\left(\boldsymbol{\hat{h}\cdot\sigma}\right)\bar{u}_{+}\right\} \right]_{\zeta=0}\label{eq:ja-1}
\end{multline}
where $\tilde{s}=1/2\cos\phi$, $\boldsymbol{\hat{h}}=\boldsymbol{h}/h$,
$\xi_{f}=\hbar v_{F}/h$ the ferromagnetic coherence length, and we
defined $u_{\pm}\equiv u_{\pm}\left(\tilde{s}\right)$ and $\bar{u}_{\pm}\equiv u_{\pm}\left(-\tilde{s}\right)$
for the sake of compactness. The presence of the $\sigma_{3}$ matrix
in \eqref{eq:ja-1} is due to the chirality in \eqref{eq:g-s-zeta}.
The anomalous current \eqref{eq:ja-1} vanishes either for a vanishing
exchange $h\rightarrow0$ or a spin-orbit $v_{\text{so}}\rightarrow0$
interaction. Since the spin-orbit effect induces a chiral propagation
along the junction, whereas the magnetic interaction induces an anisotropic
Fermi surface (see Fig.\ref{fig:Fermi-surface}) in addition to the
breaking of the time-reversal symmetry, these two conditions are minimal
for the existence of the anomalous current.

To obtain explicit values for $j_{a}$, one has to solve \eqref{eq:u_pm}.
This can be done in perturbation, supposing $h\gg p_{F}v_{\text{so}}$.
Then we have 
\begin{multline}
u_{\pm}\left(s\right)=e^{\mp\mathbf{i}L\boldsymbol{\hat{h}\cdot\sigma}s/\xi_{f}}\times\\
\left(1-\dfrac{\mathbf{i}L}{\hbar v_{F}}\int_{0}^{s}e^{\pm\mathbf{i}L\boldsymbol{\hat{h}\cdot\sigma}s_{1}/\xi_{f}}h_{A}\left(s_{1}\right)e^{\mp\mathbf{i}L\boldsymbol{\hat{h}\cdot\sigma}s_{1}/\xi_{f}}ds_{1}\right)\label{eq:u-perturb}
\end{multline}
at first order in $L/\xi_{\text{so}}$. We verified that when $v_{\text{so}}/v_{F}=0$,
then $j_{a}=0$ and $j_{c}\propto j_{\Delta}\cos\left(2L/\xi_{f}\right)$,
as usual for a pure S/F/S system \cite{buzdin.2005_RMP}, which further
reduces to the S/N/S case when the exchange field disappears \cite{golubov_kupriyanov.2004}.
Also, even for $v_{\text{so}}/v_{F}\neq0$ but when $\boldsymbol{h}$
is along the $z$-axis, then $j_{a}=0$ in \eqref{eq:ja-1}. So
to break the time-reversal symmetry in a chiral system is not sufficient
to obtain $j_{a}\neq0$: the anisotropy of the Fermi surface is also
mandatory (see Fig.\ref{fig:Fermi-surface}).

We finally inject \eqref{eq:u-perturb} into \eqref{eq:ja-1} and
\eqref{eq:jc} and we obtain
\begin{equation}
\dfrac{j_{c}}{j_{\Delta}}=-\sqrt{8\pi}\dfrac{e^{-L/2\xi_{T}}}{\sqrt{L/2\xi_{T}}}\cos\dfrac{2L}{\xi_{f}}\label{eq:jc-ST}
\end{equation}
\begin{equation}
\dfrac{j_{a}}{j_{\Delta}}=\sqrt{2\pi}\dfrac{L^{3}}{\xi_{f}\xi_{\text{so}}^{2}}\dfrac{v_{\text{so}}}{v_{F}}\sin\theta\dfrac{e^{-L/2\xi_{T}}}{\left(L/2\xi_{T}\right)^{3/2}}\sin\dfrac{L}{\xi_{f}}\label{eq:ja-ST}
\end{equation}
up to the first non-trivial term in the approximation $h\gg p_{F}v_{\text{so}}$.
Note that \eqref{eq:CPR-clean} can also be written $j\approx j_{c}\sin\left(\varphi-\varphi_{0}\right)$
at the first non-trivial order in $v_{\text{so}}/v_{F}$, with a magneto-electric
phase-shift
\begin{equation}
\tan\varphi_{0}=\dfrac{1}{2}\dfrac{L^{2}\xi_{T}}{\xi_{\text{so}}^{2}\xi_{f}}\dfrac{v_{\text{so}}}{v_{F}}\sin\theta\dfrac{\sin L/\xi_{f}}{\cos2L/\xi_{f}}\label{eq:phi-0}
\end{equation}
and thus presents some phase-kinks when the cosine term in \eqref{eq:jc-ST}
vanishes. These phase-kink are reminiscent of the transition between
the $0$- and $\pi$-phases in S/F/S-JJ \cite{buzdin.2005_RMP}. 

A relation similar to \eqref{eq:phi-0} was obtained by Buzdin
in the case of an interacting fermionic gas, see \cite{buzdin:107005.2008}
and discussion after Eq.\eqref{eq:H}. We here recover his result
in the case of a simpler Fermi liquid with non-trivial in-plane magnetic
texture.

The current-phase relation \eqref{eq:CPR-clean} is remarkable. First,
the critical-current $j_{c}$ in \eqref{eq:jc-ST} exhibits oscillations
with respect to the length of the junction, as usual with S/F proximity
effect \cite{buzdin.2005_RMP}. In addition, a SQUID made with a S/ST/S-JJ
exhibits a self-generated flux \cite{Mints1998,buzdin:107005.2008},
due to the anomalous current in \eqref{eq:ja-ST}. This flux can
be quenched by changing the relative orientation of the exchange field
with respect to the direction of the junction (the $\sin\theta$ term
in \eqref{eq:ja-ST}), \textit{e.g.} by the application of an
external magnetic field in the plane of the junction. Also, the spin-orbit
interaction $v_{\text{so}}$ can be modified by the application of
a gate-voltage, as demonstrated for 2D-JJ \cite{Takayanagi1985,Kresin1986},
which in \eqref{eq:CPR-clean} also changes the strength of the
anomalous current $j_{a}$, proportional to $\left(v_{\text{so}}/v_{F}\right)^{3}$
in the limit $h\gg p_{F}v_{\text{so}}$. All these effects can be
easily demonstrated using currently available nano-technologies. Note
also the possibility to excite the magnetization using a voltage-biased
S/ST/S-JJ, as shown in \cite{Konschelle2009}.

It is delicate to compare our results with the numerical ones \cite{reynoso_etal:107001.2008,Reynoso2012,Zazunov2009,Liu2010,Yokoyama2012,Yokoyama2014},
since our quasi-classic treatment in principle accounts for a large
number of channels, whereas the numerical works focused essentially
on a small number of channels. Note the exception \cite{Reynoso2012},
where an almost sinusoidal current-phase relation has been found numerically
in the many-channels situation, in a form similar to \eqref{eq:CPR-clean}.
Also, to have chiral propagation has been widely acknowledged as a
necessary condition for the obtention of a $\varphi_{0}$-junction
\cite{reynoso_etal:107001.2008,Reynoso2012,Zazunov2009,Yokoyama2012,Yokoyama2014}.
Moreover the dependency of $j_{a}$ with respect to the orientation
of the magnetic effect has been discussed in a few occasions \cite{reynoso_etal:107001.2008,Zazunov2009,Liu2010,Reynoso2012,Yokoyama2012}.
I hope that the present study may help understanding the geometric
nature of the magneto-electric coupling in the anomalous current \eqref{eq:ja-1}.
Also, I clearly showed that the presence of the quantum-point-contact
is not a requirement for the anomalous current to exist (see also
\cite{krive_kadigrobov_shekhter_jonson.2005,buzdin:107005.2008,Liu2010,Reynoso2012}).
In addition, the $\varphi_{0}$-behavior does not require neither
a long junction nor an interacting gas to exist.

In conclusion, using a gauge-covariant transport formalism to take
into account the spin and charge degrees of freedom of the Cooper
pairs on equal footing, I showed that a magneto-electric coupling
arrises in spin-textured Josephson junctions. This magneto-electric
effect is due to the chirality of the propagation modes inside the
spin-textured region, its explicit geometric nature being due to the
anisotropy of the Fermi surface induced by the breaking of the time-reversal-symmetry,
see Fig. \ref{fig:Fermi-surface}. More importantly, these two criteria
(anisotropic chirality and time-reversal-symmetry breaking) are the
minimal requirements for the existence of the magneto-electric coupling.
I here established an anomalous current-phase relation $j=j_{0}\sin\left(\varphi-\varphi_{0}\right)$
(\textit{i.e.} $j=j_{c}\sin\varphi+j_{a}\cos\varphi$, Eqs. \eqref{eq:CPR-clean}-\eqref{eq:ja})
in the proximity with the critical temperature of a ballistic junction,
with $\varphi_{0}\propto\boldsymbol{\left(h\times n\right)\cdot\nabla\varphi}$
a geometric term coupling the exchange field $\boldsymbol{h}$ to
the superconducting phase difference $\varphi$ via the spin-orbit
orientation $\boldsymbol{n}$, according to \eqref{eq:phi-0}.
Such a $\varphi_{0}$-phase-shift induces an anomalous current $j_{a}\neq0$,
Eq.\eqref{eq:ja-ST}. Note finally that similar results exist in the
diffusive limit %
\footnote{F.S. Bergeret and I.V. Tokatly, \textit{in preparation}.%
} and that a more detailed version of this work is in preparation.
\begin{acknowledgments}
I thank F. Hassler and G. Viola for daily stimulating discussions.
Special thanks are due to A.I. Buzdin for his patient introduction
to the physics of $\varphi_{0}$-junction. This work also benefited
from enlightening discussions with F.S. Bergeret, A. Larat and I.V.
Tokatly. I am grateful for support from the Alexander von Humboldt
foundation.
\end{acknowledgments}
\bibliographystyle{apsrev4-1}
\bibliography{letter}

\begin{thebibliography}{32}%
\makeatletter
\providecommand \@ifxundefined [1]{%
 \@ifx{#1\undefined}
}%
\providecommand \@ifnum [1]{%
 \ifnum #1\expandafter \@firstoftwo
 \else \expandafter \@secondoftwo
 \fi
}%
\providecommand \@ifx [1]{%
 \ifx #1\expandafter \@firstoftwo
 \else \expandafter \@secondoftwo
 \fi
}%
\providecommand \natexlab [1]{#1}%
\providecommand \enquote  [1]{``#1''}%
\providecommand \bibnamefont  [1]{#1}%
\providecommand \bibfnamefont [1]{#1}%
\providecommand \citenamefont [1]{#1}%
\providecommand \href@noop [0]{\@secondoftwo}%
\providecommand \href [0]{\begingroup \@sanitize@url \@href}%
\providecommand \@href[1]{\@@startlink{#1}\@@href}%
\providecommand \@@href[1]{\endgroup#1\@@endlink}%
\providecommand \@sanitize@url [0]{\catcode `\\12\catcode `\$12\catcode
  `\&12\catcode `\#12\catcode `\^12\catcode `\_12\catcode `\%12\relax}%
\providecommand \@@startlink[1]{}%
\providecommand \@@endlink[0]{}%
\providecommand \url  [0]{\begingroup\@sanitize@url \@url }%
\providecommand \@url [1]{\endgroup\@href {#1}{\urlprefix }}%
\providecommand \urlprefix  [0]{URL }%
\providecommand \Eprint [0]{\href }%
\providecommand \doibase [0]{http://dx.doi.org/}%
\providecommand \selectlanguage [0]{\@gobble}%
\providecommand \bibinfo  [0]{\@secondoftwo}%
\providecommand \bibfield  [0]{\@secondoftwo}%
\providecommand \translation [1]{[#1]}%
\providecommand \BibitemOpen [0]{}%
\providecommand \bibitemStop [0]{}%
\providecommand \bibitemNoStop [0]{.\EOS\space}%
\providecommand \EOS [0]{\spacefactor3000\relax}%
\providecommand \BibitemShut  [1]{\csname bibitem#1\endcsname}%
\let\auto@bib@innerbib\@empty
%</preamble>
\bibitem [{\citenamefont {Doh}\ \emph {et~al.}(2005)\citenamefont {Doh},
  \citenamefont {van Dam}, \citenamefont {Roest}, \citenamefont {Bakkers},
  \citenamefont {Kouwenhoven},\ and\ \citenamefont {{De
  Franceschi}}}]{Doh2005}%
  \BibitemOpen
  \bibfield  {author} {\bibinfo {author} {\bibfnamefont {Y.-J.}\ \bibnamefont
  {Doh}}, \bibinfo {author} {\bibfnamefont {J.~A.}\ \bibnamefont {van Dam}},
  \bibinfo {author} {\bibfnamefont {A.~L.}\ \bibnamefont {Roest}}, \bibinfo
  {author} {\bibfnamefont {E.~P. a.~M.}\ \bibnamefont {Bakkers}}, \bibinfo
  {author} {\bibfnamefont {L.~P.}\ \bibnamefont {Kouwenhoven}}, \ and\ \bibinfo
  {author} {\bibfnamefont {S.}~\bibnamefont {{De Franceschi}}},\ }\href
  {\doibase 10.1126/science.1113523} {\bibfield  {journal} {\bibinfo  {journal}
  {Science (New York, N.Y.)}\ }\textbf {\bibinfo {volume} {309}},\ \bibinfo
  {pages} {272} (\bibinfo {year} {2005})},\ \Eprint
  {http://arxiv.org/abs/0508558} {arXiv:0508558 [cond-mat]} \BibitemShut
  {NoStop}%
\bibitem [{\citenamefont {Xiang}\ \emph {et~al.}(2006)\citenamefont {Xiang},
  \citenamefont {Vidan}, \citenamefont {Tinkham}, \citenamefont {Westervelt},\
  and\ \citenamefont {Lieber}}]{Xiang2006}%
  \BibitemOpen
  \bibfield  {author} {\bibinfo {author} {\bibfnamefont {J.}~\bibnamefont
  {Xiang}}, \bibinfo {author} {\bibfnamefont {A.}~\bibnamefont {Vidan}},
  \bibinfo {author} {\bibfnamefont {M.}~\bibnamefont {Tinkham}}, \bibinfo
  {author} {\bibfnamefont {R.~M.}\ \bibnamefont {Westervelt}}, \ and\ \bibinfo
  {author} {\bibfnamefont {C.~M.}\ \bibnamefont {Lieber}},\ }\href {\doibase
  10.1038/nnano.2006.140} {\bibfield  {journal} {\bibinfo  {journal} {Nature
  nanotechnology}\ }\textbf {\bibinfo {volume} {1}},\ \bibinfo {pages} {208}
  (\bibinfo {year} {2006})}\BibitemShut {NoStop}%
\bibitem [{\citenamefont {Nilsson}\ \emph {et~al.}(2012)\citenamefont
  {Nilsson}, \citenamefont {Samuelsson}, \citenamefont {Caroff},\ and\
  \citenamefont {Xu}}]{Nilsson2012}%
  \BibitemOpen
  \bibfield  {author} {\bibinfo {author} {\bibfnamefont {H.~A.}\ \bibnamefont
  {Nilsson}}, \bibinfo {author} {\bibfnamefont {P.}~\bibnamefont {Samuelsson}},
  \bibinfo {author} {\bibfnamefont {P.}~\bibnamefont {Caroff}}, \ and\ \bibinfo
  {author} {\bibfnamefont {H.~Q.}\ \bibnamefont {Xu}},\ }\href {\doibase
  10.1021/nl203380w} {\bibfield  {journal} {\bibinfo  {journal} {Nano letters}\
  }\textbf {\bibinfo {volume} {12}},\ \bibinfo {pages} {228} (\bibinfo {year}
  {2012})},\ \Eprint {http://arxiv.org/abs/1204.3936} {arXiv:1204.3936}
  \BibitemShut {NoStop}%
\bibitem [{\citenamefont {Buzdin}(2005{\natexlab{a}})}]{buzdin.2005_RMP}%
  \BibitemOpen
  \bibfield  {author} {\bibinfo {author} {\bibfnamefont {A.~I.}\ \bibnamefont
  {Buzdin}},\ }\href {\doibase 10.1103/RevModPhys.77.935} {\bibfield  {journal}
  {\bibinfo  {journal} {Reviews of Modern Physics}\ }\textbf {\bibinfo {volume}
  {77}},\ \bibinfo {pages} {935} (\bibinfo {year}
  {2005}{\natexlab{a}})}\BibitemShut {NoStop}%
\bibitem [{\citenamefont {Bergeret}\ \emph {et~al.}(2005)\citenamefont
  {Bergeret}, \citenamefont {Volkov},\ and\ \citenamefont
  {Efetov}}]{bergeret_volkov_efetov_R.2005}%
  \BibitemOpen
  \bibfield  {author} {\bibinfo {author} {\bibfnamefont {F.}~\bibnamefont
  {Bergeret}}, \bibinfo {author} {\bibfnamefont {A.~F.}\ \bibnamefont
  {Volkov}}, \ and\ \bibinfo {author} {\bibfnamefont {K.~B.}\ \bibnamefont
  {Efetov}},\ }\href {\doibase 10.1103/RevModPhys.77.1321} {\bibfield
  {journal} {\bibinfo  {journal} {Reviews of Modern Physics}\ }\textbf
  {\bibinfo {volume} {77}},\ \bibinfo {pages} {1321} (\bibinfo {year}
  {2005})}\BibitemShut {NoStop}%
\bibitem [{\citenamefont {Casalbuoni}\ and\ \citenamefont
  {Nardulli}(2004)}]{casalbuoni.nardulli.2004}%
  \BibitemOpen
  \bibfield  {author} {\bibinfo {author} {\bibfnamefont {R.}~\bibnamefont
  {Casalbuoni}}\ and\ \bibinfo {author} {\bibfnamefont {G.}~\bibnamefont
  {Nardulli}},\ }\href@noop {} {\bibfield  {journal} {\bibinfo  {journal}
  {Reviews of Modern Physics}\ }\textbf {\bibinfo {volume} {76}},\ \bibinfo
  {pages} {263} (\bibinfo {year} {2004})}\BibitemShut {NoStop}%
\bibitem [{\citenamefont {Alford}\ \emph {et~al.}(2008)\citenamefont {Alford},
  \citenamefont {Schmitt}, \citenamefont {Rajagopal},\ and\ \citenamefont
  {Sch\"{a}fer}}]{Alford2008}%
  \BibitemOpen
  \bibfield  {author} {\bibinfo {author} {\bibfnamefont {M.}~\bibnamefont
  {Alford}}, \bibinfo {author} {\bibfnamefont {A.}~\bibnamefont {Schmitt}},
  \bibinfo {author} {\bibfnamefont {K.}~\bibnamefont {Rajagopal}}, \ and\
  \bibinfo {author} {\bibfnamefont {T.}~\bibnamefont {Sch\"{a}fer}},\ }\href
  {\doibase 10.1103/RevModPhys.80.1455} {\bibfield  {journal} {\bibinfo
  {journal} {Reviews of Modern Physics}\ }\textbf {\bibinfo {volume} {80}},\
  \bibinfo {pages} {1455} (\bibinfo {year} {2008})},\ \Eprint
  {http://arxiv.org/abs/0709.4635} {arXiv:0709.4635} \BibitemShut {NoStop}%
\bibitem [{\citenamefont {Nayak}\ \emph {et~al.}(2008)\citenamefont {Nayak},
  \citenamefont {Stern}, \citenamefont {Freedman},\ and\ \citenamefont {{Das
  Sarma}}}]{Nayak2008}%
  \BibitemOpen
  \bibfield  {author} {\bibinfo {author} {\bibfnamefont {C.}~\bibnamefont
  {Nayak}}, \bibinfo {author} {\bibfnamefont {A.}~\bibnamefont {Stern}},
  \bibinfo {author} {\bibfnamefont {M.~H.}\ \bibnamefont {Freedman}}, \ and\
  \bibinfo {author} {\bibfnamefont {S.}~\bibnamefont {{Das Sarma}}},\ }\href
  {\doibase 10.1103/RevModPhys.80.1083} {\bibfield  {journal} {\bibinfo
  {journal} {Reviews of Modern Physics}\ }\textbf {\bibinfo {volume} {80}},\
  \bibinfo {pages} {1083} (\bibinfo {year} {2008})},\ \Eprint
  {http://arxiv.org/abs/0707.1889} {arXiv:0707.1889} \BibitemShut {NoStop}%
\bibitem [{\citenamefont {Georgescu}\ \emph {et~al.}(2014)\citenamefont
  {Georgescu}, \citenamefont {Ashhab},\ and\ \citenamefont
  {Nori}}]{Georgescu2014}%
  \BibitemOpen
  \bibfield  {author} {\bibinfo {author} {\bibfnamefont {I.~M.}\ \bibnamefont
  {Georgescu}}, \bibinfo {author} {\bibfnamefont {S.}~\bibnamefont {Ashhab}}, \
  and\ \bibinfo {author} {\bibfnamefont {F.}~\bibnamefont {Nori}},\ }\href
  {\doibase 10.1103/RevModPhys.86.153} {\bibfield  {journal} {\bibinfo
  {journal} {Reviews of Modern Physics}\ }\textbf {\bibinfo {volume} {86}},\
  \bibinfo {pages} {153} (\bibinfo {year} {2014})},\ \Eprint
  {http://arxiv.org/abs/1308.6253} {arXiv:1308.6253} \BibitemShut {NoStop}%
\bibitem [{\citenamefont {Mints}(1998)}]{Mints1998}%
  \BibitemOpen
  \bibfield  {author} {\bibinfo {author} {\bibfnamefont {R.}~\bibnamefont
  {Mints}},\ }\href {\doibase 10.1103/PhysRevB.57.R3221} {\bibfield  {journal}
  {\bibinfo  {journal} {Physical Review B}\ }\textbf {\bibinfo {volume} {57}},\
  \bibinfo {pages} {R3221} (\bibinfo {year} {1998})}\BibitemShut {NoStop}%
\bibitem [{\citenamefont {Buzdin}(2005{\natexlab{b}})}]{buzdin.2005}%
  \BibitemOpen
  \bibfield  {author} {\bibinfo {author} {\bibfnamefont {A.~I.}\ \bibnamefont
  {Buzdin}},\ }\href {\doibase 10.1103/PhysRevB.72.100501} {\bibfield
  {journal} {\bibinfo  {journal} {Physical Review B}\ }\textbf {\bibinfo
  {volume} {72}},\ \bibinfo {pages} {100501} (\bibinfo {year}
  {2005}{\natexlab{b}})}\BibitemShut {NoStop}%
\bibitem [{\citenamefont {Sickinger}\ \emph {et~al.}(2012)\citenamefont
  {Sickinger}, \citenamefont {Lipman}, \citenamefont {Weides}, \citenamefont
  {Mints}, \citenamefont {Kohlstedt}, \citenamefont {Koelle}, \citenamefont
  {Kleiner},\ and\ \citenamefont {Goldobin}}]{Sickinger2012}%
  \BibitemOpen
  \bibfield  {author} {\bibinfo {author} {\bibfnamefont {H.}~\bibnamefont
  {Sickinger}}, \bibinfo {author} {\bibfnamefont {A.}~\bibnamefont {Lipman}},
  \bibinfo {author} {\bibfnamefont {M.}~\bibnamefont {Weides}}, \bibinfo
  {author} {\bibfnamefont {R.~G.}\ \bibnamefont {Mints}}, \bibinfo {author}
  {\bibfnamefont {H.}~\bibnamefont {Kohlstedt}}, \bibinfo {author}
  {\bibfnamefont {D.}~\bibnamefont {Koelle}}, \bibinfo {author} {\bibfnamefont
  {R.}~\bibnamefont {Kleiner}}, \ and\ \bibinfo {author} {\bibfnamefont
  {E.}~\bibnamefont {Goldobin}},\ }\href {\doibase
  10.1103/PhysRevLett.109.107002} {\bibfield  {journal} {\bibinfo  {journal}
  {Physical Review Letters}\ }\textbf {\bibinfo {volume} {109}},\ \bibinfo
  {pages} {107002} (\bibinfo {year} {2012})},\ \Eprint
  {http://arxiv.org/abs/1207.3013} {arXiv:1207.3013} \BibitemShut {NoStop}%
\bibitem [{\citenamefont {Krive}\ \emph {et~al.}(2004)\citenamefont {Krive},
  \citenamefont {Kadigrobov}, \citenamefont {Shekhter},\ and\ \citenamefont
  {Jonson}}]{krive_kadigrobov_shekhter_jonson.2005}%
  \BibitemOpen
  \bibfield  {author} {\bibinfo {author} {\bibfnamefont {I.~V.}\ \bibnamefont
  {Krive}}, \bibinfo {author} {\bibfnamefont {A.~M.}\ \bibnamefont
  {Kadigrobov}}, \bibinfo {author} {\bibfnamefont {R.~I.}\ \bibnamefont
  {Shekhter}}, \ and\ \bibinfo {author} {\bibfnamefont {M.}~\bibnamefont
  {Jonson}},\ }\href {\doibase 10.1103/PhysRevB.71.214516} {\bibfield
  {journal} {\bibinfo  {journal} {Physical Review B}\ }\textbf {\bibinfo
  {volume} {71}},\ \bibinfo {pages} {214516} (\bibinfo {year} {2004})},\
  \Eprint {http://arxiv.org/abs/0409063} {arXiv:0409063 [cond-mat]}
  \BibitemShut {NoStop}%
\bibitem [{\citenamefont {Buzdin}(2008)}]{buzdin:107005.2008}%
  \BibitemOpen
  \bibfield  {author} {\bibinfo {author} {\bibfnamefont {A.~I.}\ \bibnamefont
  {Buzdin}},\ }\href {\doibase 10.1103/PhysRevLett.101.107005} {\bibfield
  {journal} {\bibinfo  {journal} {Physical Review Letters}\ }\textbf {\bibinfo
  {volume} {101}},\ \bibinfo {pages} {107005} (\bibinfo {year} {2008})},\
  \Eprint {http://arxiv.org/abs/0808.0299} {arXiv:0808.0299} \BibitemShut
  {NoStop}%
\bibitem [{\citenamefont {Reynoso}\ \emph {et~al.}(2008)\citenamefont
  {Reynoso}, \citenamefont {Usaj}, \citenamefont {Balseiro}, \citenamefont
  {Feinberg},\ and\ \citenamefont {Avignon}}]{reynoso_etal:107001.2008}%
  \BibitemOpen
  \bibfield  {author} {\bibinfo {author} {\bibfnamefont {A.~A.}\ \bibnamefont
  {Reynoso}}, \bibinfo {author} {\bibfnamefont {G.}~\bibnamefont {Usaj}},
  \bibinfo {author} {\bibfnamefont {C.}~\bibnamefont {Balseiro}}, \bibinfo
  {author} {\bibfnamefont {D.}~\bibnamefont {Feinberg}}, \ and\ \bibinfo
  {author} {\bibfnamefont {M.}~\bibnamefont {Avignon}},\ }\href {\doibase
  10.1103/PhysRevLett.101.107001} {\bibfield  {journal} {\bibinfo  {journal}
  {Physical Review Letters}\ }\textbf {\bibinfo {volume} {101}},\ \bibinfo
  {pages} {107001} (\bibinfo {year} {2008})},\ \Eprint
  {http://arxiv.org/abs/0808.1516} {arXiv:0808.1516} \BibitemShut {NoStop}%
\bibitem [{\citenamefont {Zazunov}\ \emph {et~al.}(2009)\citenamefont
  {Zazunov}, \citenamefont {Egger}, \citenamefont {Martin},\ and\ \citenamefont
  {Jonckheere}}]{Zazunov2009}%
  \BibitemOpen
  \bibfield  {author} {\bibinfo {author} {\bibfnamefont {A.}~\bibnamefont
  {Zazunov}}, \bibinfo {author} {\bibfnamefont {R.}~\bibnamefont {Egger}},
  \bibinfo {author} {\bibfnamefont {T.}~\bibnamefont {Martin}}, \ and\ \bibinfo
  {author} {\bibfnamefont {T.}~\bibnamefont {Jonckheere}},\ }\href {\doibase
  10.1103/PhysRevLett.103.147004} {\bibfield  {journal} {\bibinfo  {journal}
  {Physical Review Letters}\ }\textbf {\bibinfo {volume} {103}},\ \bibinfo
  {pages} {147004} (\bibinfo {year} {2009})},\ \Eprint
  {http://arxiv.org/abs/0909.3036} {arXiv:0909.3036} \BibitemShut {NoStop}%
\bibitem [{\citenamefont {Liu}\ and\ \citenamefont {Chan}(2010)}]{Liu2010}%
  \BibitemOpen
  \bibfield  {author} {\bibinfo {author} {\bibfnamefont {J.-F.}\ \bibnamefont
  {Liu}}\ and\ \bibinfo {author} {\bibfnamefont {K.}~\bibnamefont {Chan}},\
  }\href {\doibase 10.1103/PhysRevB.82.125305} {\bibfield  {journal} {\bibinfo
  {journal} {Physical Review B}\ }\textbf {\bibinfo {volume} {82}},\ \bibinfo
  {pages} {125305} (\bibinfo {year} {2010})}\BibitemShut {NoStop}%
\bibitem [{\citenamefont {Reynoso}\ \emph {et~al.}(2012)\citenamefont
  {Reynoso}, \citenamefont {Usaj}, \citenamefont {Balseiro}, \citenamefont
  {Feinberg},\ and\ \citenamefont {Avignon}}]{Reynoso2012}%
  \BibitemOpen
  \bibfield  {author} {\bibinfo {author} {\bibfnamefont {A.~A.}\ \bibnamefont
  {Reynoso}}, \bibinfo {author} {\bibfnamefont {G.}~\bibnamefont {Usaj}},
  \bibinfo {author} {\bibfnamefont {C.~A.}\ \bibnamefont {Balseiro}}, \bibinfo
  {author} {\bibfnamefont {D.}~\bibnamefont {Feinberg}}, \ and\ \bibinfo
  {author} {\bibfnamefont {M.}~\bibnamefont {Avignon}},\ }\href {\doibase
  10.1103/PhysRevB.86.214519} {\bibfield  {journal} {\bibinfo  {journal}
  {Physical Review B}\ }\textbf {\bibinfo {volume} {86}},\ \bibinfo {pages}
  {214519} (\bibinfo {year} {2012})},\ \Eprint {http://arxiv.org/abs/1212.2786}
  {arXiv:1212.2786} \BibitemShut {NoStop}%
\bibitem [{\citenamefont {Yokoyama}\ \emph {et~al.}(2013)\citenamefont
  {Yokoyama}, \citenamefont {Eto},\ and\ \citenamefont {{V.
  Nazarov}}}]{Yokoyama2012}%
  \BibitemOpen
  \bibfield  {author} {\bibinfo {author} {\bibfnamefont {T.}~\bibnamefont
  {Yokoyama}}, \bibinfo {author} {\bibfnamefont {M.}~\bibnamefont {Eto}}, \
  and\ \bibinfo {author} {\bibfnamefont {Y.}~\bibnamefont {{V. Nazarov}}},\
  }\href {\doibase 10.7566/JPSJ.82.054703} {\bibfield  {journal} {\bibinfo
  {journal} {Journal of the Physical Society of Japan}\ }\textbf {\bibinfo
  {volume} {82}},\ \bibinfo {pages} {054703} (\bibinfo {year} {2013})},\
  \Eprint {http://arxiv.org/abs/1212.5390} {arXiv:1212.5390} \BibitemShut
  {NoStop}%
\bibitem [{\citenamefont {Yokoyama}\ \emph {et~al.}(2014)\citenamefont
  {Yokoyama}, \citenamefont {Eto},\ and\ \citenamefont
  {Nazarov}}]{Yokoyama2014}%
  \BibitemOpen
  \bibfield  {author} {\bibinfo {author} {\bibfnamefont {T.}~\bibnamefont
  {Yokoyama}}, \bibinfo {author} {\bibfnamefont {M.}~\bibnamefont {Eto}}, \
  and\ \bibinfo {author} {\bibfnamefont {Y.~V.}\ \bibnamefont {Nazarov}},\
  }\href {http://arxiv.org/abs/1402.0305} {\  (\bibinfo {year} {2014})},\
  \Eprint {http://arxiv.org/abs/1402.0305} {arXiv:1402.0305} \BibitemShut
  {NoStop}%
\bibitem [{\citenamefont {Golubov}\ \emph {et~al.}(2004)\citenamefont
  {Golubov}, \citenamefont {Kupriyanov},\ and\ \citenamefont
  {Il'ichev}}]{golubov_kupriyanov.2004}%
  \BibitemOpen
  \bibfield  {author} {\bibinfo {author} {\bibfnamefont {A.}~\bibnamefont
  {Golubov}}, \bibinfo {author} {\bibfnamefont {M.}~\bibnamefont {Kupriyanov}},
  \ and\ \bibinfo {author} {\bibfnamefont {E.}~\bibnamefont {Il'ichev}},\
  }\href {\doibase 10.1103/RevModPhys.76.411} {\bibfield  {journal} {\bibinfo
  {journal} {Reviews of Modern Physics}\ }\textbf {\bibinfo {volume} {76}},\
  \bibinfo {pages} {411} (\bibinfo {year} {2004})}\BibitemShut {NoStop}%
\bibitem [{\citenamefont {Takayanagi}\ and\ \citenamefont
  {Kawakami}(1985)}]{Takayanagi1985}%
  \BibitemOpen
  \bibfield  {author} {\bibinfo {author} {\bibfnamefont {H.}~\bibnamefont
  {Takayanagi}}\ and\ \bibinfo {author} {\bibfnamefont {T.}~\bibnamefont
  {Kawakami}},\ }\href {\doibase 10.1103/PhysRevLett.54.2449} {\bibfield
  {journal} {\bibinfo  {journal} {Physical Review Letters}\ }\textbf {\bibinfo
  {volume} {54}},\ \bibinfo {pages} {2449} (\bibinfo {year}
  {1985})}\BibitemShut {NoStop}%
\bibitem [{\citenamefont {Kresin}(1986)}]{Kresin1986}%
  \BibitemOpen
  \bibfield  {author} {\bibinfo {author} {\bibfnamefont {V.}~\bibnamefont
  {Kresin}},\ }\href {\doibase 10.1103/PhysRevB.34.7587} {\bibfield  {journal}
  {\bibinfo  {journal} {Physical Review B}\ }\textbf {\bibinfo {volume} {34}},\
  \bibinfo {pages} {7587} (\bibinfo {year} {1986})}\BibitemShut {NoStop}%
\bibitem [{\citenamefont {Edelstein}(1996)}]{Edelstein1996}%
  \BibitemOpen
  \bibfield  {author} {\bibinfo {author} {\bibfnamefont {V.~M.}\ \bibnamefont
  {Edelstein}},\ }\href {\doibase 10.1088/0953-8984/8/3/012} {\bibfield
  {journal} {\bibinfo  {journal} {Journal of Physics: Condensed Matter}\
  }\textbf {\bibinfo {volume} {8}},\ \bibinfo {pages} {339} (\bibinfo {year}
  {1996})}\BibitemShut {NoStop}%
\bibitem [{\citenamefont {Konschelle}\ and\ \citenamefont
  {Buzdin}(2009)}]{Konschelle2009}%
  \BibitemOpen
  \bibfield  {author} {\bibinfo {author} {\bibfnamefont {F.}~\bibnamefont
  {Konschelle}}\ and\ \bibinfo {author} {\bibfnamefont {A.~I.}\ \bibnamefont
  {Buzdin}},\ }\href {\doibase 10.1103/PhysRevLett.102.017001} {\bibfield
  {journal} {\bibinfo  {journal} {Physical Review Letters}\ }\textbf {\bibinfo
  {volume} {102}},\ \bibinfo {pages} {017001} (\bibinfo {year} {2009})},\
  \Eprint {http://arxiv.org/abs/0810.4286} {arXiv:0810.4286} \BibitemShut
  {NoStop}%
\bibitem [{\citenamefont {Bergeret}\ and\ \citenamefont
  {Tokatly}(2014)}]{Bergeret2014}%
  \BibitemOpen
  \bibfield  {author} {\bibinfo {author} {\bibfnamefont {F.~S.}\ \bibnamefont
  {Bergeret}}\ and\ \bibinfo {author} {\bibfnamefont {I.~V.}\ \bibnamefont
  {Tokatly}},\ }\href {\doibase 10.1103/PhysRevB.89.134517} {\bibfield
  {journal} {\bibinfo  {journal} {Physical Review B}\ }\textbf {\bibinfo
  {volume} {89}},\ \bibinfo {pages} {134517} (\bibinfo {year} {2014})},\
  \Eprint {http://arxiv.org/abs/1402.1025} {arXiv:1402.1025} \BibitemShut
  {NoStop}%
\bibitem [{\citenamefont {Konschelle}(2014)}]{Konschelle2014}%
  \BibitemOpen
  \bibfield  {author} {\bibinfo {author} {\bibfnamefont {F.}~\bibnamefont
  {Konschelle}},\ }\href {\doibase 10.1140/epjb/e2014-50143-0} {\bibfield
  {journal} {\bibinfo  {journal} {The European Physical Journal B}\ }\textbf
  {\bibinfo {volume} {87}},\ \bibinfo {pages} {119} (\bibinfo {year} {2014})},\
  \Eprint {http://arxiv.org/abs/1403.1797} {arXiv:1403.1797} \BibitemShut
  {NoStop}%
\bibitem [{\citenamefont {Cheng}\ and\ \citenamefont
  {Lutchyn}(2012)}]{Cheng2012a}%
  \BibitemOpen
  \bibfield  {author} {\bibinfo {author} {\bibfnamefont {M.}~\bibnamefont
  {Cheng}}\ and\ \bibinfo {author} {\bibfnamefont {R.~M.}\ \bibnamefont
  {Lutchyn}},\ }\href {\doibase 10.1103/PhysRevB.86.134522} {\bibfield
  {journal} {\bibinfo  {journal} {Physical Review B}\ }\textbf {\bibinfo
  {volume} {86}},\ \bibinfo {pages} {134522} (\bibinfo {year} {2012})},\
  \Eprint {http://arxiv.org/abs/1201.1918} {arXiv:1201.1918} \BibitemShut
  {NoStop}%
\bibitem [{\citenamefont {Rokhinson}\ \emph {et~al.}(2012)\citenamefont
  {Rokhinson}, \citenamefont {Liu},\ and\ \citenamefont
  {Furdyna}}]{Rokhinson2012}%
  \BibitemOpen
  \bibfield  {author} {\bibinfo {author} {\bibfnamefont {L.~P.}\ \bibnamefont
  {Rokhinson}}, \bibinfo {author} {\bibfnamefont {X.}~\bibnamefont {Liu}}, \
  and\ \bibinfo {author} {\bibfnamefont {J.~K.}\ \bibnamefont {Furdyna}},\
  }\href {\doibase 10.1038/nphys2429} {\bibfield  {journal} {\bibinfo
  {journal} {Nature Physics}\ }\textbf {\bibinfo {volume} {8}},\ \bibinfo
  {pages} {795} (\bibinfo {year} {2012})},\ \Eprint
  {http://arxiv.org/abs/1204.4212} {arXiv:1204.4212} \BibitemShut {NoStop}%
\bibitem [{\citenamefont {Schopohl}\ and\ \citenamefont
  {Maki}(1995)}]{Schopohl1995}%
  \BibitemOpen
  \bibfield  {author} {\bibinfo {author} {\bibfnamefont {N.}~\bibnamefont
  {Schopohl}}\ and\ \bibinfo {author} {\bibfnamefont {K.}~\bibnamefont
  {Maki}},\ }\href {\doibase 10.1103/PhysRevB.52.490} {\bibfield  {journal}
  {\bibinfo  {journal} {Physical Review B}\ }\textbf {\bibinfo {volume} {52}},\
  \bibinfo {pages} {490} (\bibinfo {year} {1995})}\BibitemShut {NoStop}%
\bibitem [{\citenamefont {Schopohl}(1998)}]{Schopohl1998}%
  \BibitemOpen
  \bibfield  {author} {\bibinfo {author} {\bibfnamefont {N.}~\bibnamefont
  {Schopohl}},\ }\href {http://arxiv.org/abs/cond-mat/9804064} {\  (\bibinfo
  {year} {1998})},\ \Eprint {http://arxiv.org/abs/9804064} {arXiv:9804064
  [cond-mat]} \BibitemShut {NoStop}%
\bibitem [{\citenamefont {Blanes}\ \emph {et~al.}(2009)\citenamefont {Blanes},
  \citenamefont {Casas}, \citenamefont {Oteo},\ and\ \citenamefont
  {Ros}}]{Blanes2009}%
  \BibitemOpen
  \bibfield  {author} {\bibinfo {author} {\bibfnamefont {S.}~\bibnamefont
  {Blanes}}, \bibinfo {author} {\bibfnamefont {F.}~\bibnamefont {Casas}},
  \bibinfo {author} {\bibfnamefont {J.}~\bibnamefont {Oteo}}, \ and\ \bibinfo
  {author} {\bibfnamefont {J.}~\bibnamefont {Ros}},\ }\href {\doibase
  10.1016/j.physrep.2008.11.001} {\bibfield  {journal} {\bibinfo  {journal}
  {Physics Reports}\ }\textbf {\bibinfo {volume} {470}},\ \bibinfo {pages}
  {151} (\bibinfo {year} {2009})},\ \Eprint {http://arxiv.org/abs/0810.5488}
  {arXiv:0810.5488} \BibitemShut {NoStop}%
\end{thebibliography}%

\end{document}